\documentclass{article}
\usepackage{spconf}
\usepackage{algorithm,graphicx}
\usepackage{algorithmic}
\usepackage{cite}
\usepackage{amsfonts}
\usepackage{fancyhdr}
\usepackage{listings}%
\usepackage{arydshln}
\usepackage{amsmath,amssymb,amsthm}%
\usepackage{graphicx,psfrag,caption,subcaption}%
\usepackage{graphicx}%
\usepackage{listings}%
\usepackage{arydshln}
\usepackage{subfig}%
\usepackage{algorithmic}%
\newtheorem{mylem}{Lemma}
\newtheorem{mydef}{Definition}%
\newtheorem{myrem}{Remark}%

\DeclareMathOperator*{\argmax}{arg\,max}

\title{Subsampling for Graph Power Spectrum Estimation}
\name{Sundeep Prabhakar Chepuri and Geert Leus\thanks{This work was supported by the KAUST-MIT-TUD consortium grant~{OSR-2015-Sensors-2700}. \newline\indent\indent Software to produce results of this paper can be downloaded from \tt{http://cas.et.tudelft.nl/$\sim$sundeep/sw/gpsd.zip}}}
\address{Delft University of Technology (TU Delft), The Netherlands\\
Email:~\{s.p.chepuri; g.j.t.leus\}@tudelft.nl.}
\begin{document}
\ninept
\maketitle
\begin{abstract}
In this paper we focus on subsampling stationary random processes that reside on the vertices of undirected graphs. Second-order stationary graph signals are obtained by filtering white noise and they admit a well-defined power spectrum. Estimating the graph power spectrum forms a central component of stationary graph signal processing and related inference 
tasks. We show that by sampling a significantly smaller subset of vertices and using simple least squares, we can reconstruct the power spectrum of the graph signal from the subsampled observations, without any spectral priors. In addition, a near-optimal greedy algorithm is developed to design the subsampling scheme. 
\end{abstract}
\vspace*{-4mm}
\section{Introduction}
\vspace*{-2mm}
Processing signals residing on the vertices of graphs is recently receiving a significant amount of interest for network science applications. 
In particular, generalizing as well as drawing parallels of classical time-frequency analysis tools to {\it graph data analysis} while incorporating the irregular structure on which the graph signals are defined is an emerging area of research~\cite{shuman2013Emerging,sandryhaila2014big}. 

We are interested in sampling and processing stationary graph signals, which are stochastic processes defined on graphs with
second-order statistics that are invariant similar to time series, but in the graph setting. Second order stationary graph signals have a well-defined {\it graph power spectrum}. Stationary graph signals can be generated by filtering white noise (or any other stationary graph process) and the graph power spectrum of the filtered signal will be characterized by the squared magnitude of the frequency response of the filter.  Using the idea of graph second-order stationarity, inference problems on graphs such as smoothing, prediction, and deconvolution can be solved by designing optimum (minimum mean squared error) Wiener-like filters. Although Wiener filters for graph signals can be derived similar to time-domain signals~\cite{girault2014semi}, graph power spectrum estimation forms a crucial component of such filter designs. 


In this paper, we focus on reconstructing graph second-order statistics, more specifically the graph power spectrum by observing a reduced subset of graph nodes. The fact that we are reconstructing the graph power spectrum, instead of the graph signal enables us to subsample or sparsely sample the graph signal and yet reconstruct the power spectrum of the original graph signal, even without any spectral priors (e.g., sparsity, bandlimited with known support). This is a new and different perspective as compared to subsampling for graph signal reconstruction~\cite{anis2014towards,marques2015sampling,tsitsvero2015uncertainty} and it extends the field of compressive covariance sensing~\cite{ariananda2012compressive,Romero16CCSspm} to graph settings. 
  
We present two approaches, namely, the {\it graph spectral domain} and the {\it graph vertex domain} approach, where the processing is done in the graph frequency and data domain, respectively (hence the name).  One of the results shows that with a reduced subset of $O(\sqrt{N})$ observations and using least squares, we can reconstruct the graph power spectrum of a length-$N$ graph signal, even in the absence of spectral priors. Any available spectral priors will naturally lead to a higher compression. We provide a low-complexity and near-optimal greedy algorithm for designing the sampling matrix that essentially performs node subset selection, which is a discrete combinatorial optimization problem.
\vspace*{-2mm}
\section{Background and modeling}
\vspace*{-2mm}
Throughout the paper we denote matrices (column vectors) with upper (lower) bold face letters. The $\ell_0$-(quasi) norm refers to the number of non-zero entries in ${\boldsymbol w}$, i.e., ${\|{\boldsymbol w}\|}_0 := |\{m\,: \, w_m \neq 0\}|$. 

\vspace*{-2mm}
\subsection{Graph signals}

Consider a dataset with $N$ elements, which live on an irregular structure represented by a known undirected graph 
$\mathcal{G} = (\mathcal{V},\mathcal{E})$, where the vertex set $\mathcal{V} = \{v_1,\cdots,v_N\}$ denotes the set of nodes, and the edge set $\mathcal{E}$ reveals any connection between the nodes. We refer to such datasets as {\it graph signals}.

Let us construct the adjacency matrix ${\boldsymbol A} \in \mathbb{S}^{N}$ with a nonzero $(i,j)$th entry $[{\boldsymbol A}]_{i,j}$ denoting the strength of the edge connecting the $i$th node and the $j$th node, while the entry is set to zero if no edge exists between the $i$th node and the $j$th node. The degree of the $i$th node is defined as $d_i = \sum_{j=1}^M [{\boldsymbol A}]_{i,j}$. An associated metric, the so-called graph Laplacian is defined as 
$
{\boldsymbol L} = {\boldsymbol D} - {\boldsymbol A} \, \in \, \mathbb{S}^N,
$ 
where ${\boldsymbol D} = {\rm diag}(d_1,d_2,\cdots,d_N) \in \mathbb{R}^{N\times N}$. We introduce a symmetric matrix ${\boldsymbol S} \in \mathbb{S}^{N}$, where $[{\boldsymbol S}]_{i,j}$ is nonzero only if $i=j$ or $(i,j) \in \mathcal{E}$. The sparsity pattern of ${\boldsymbol S}$ captures the local structure of the graph, hence ${\boldsymbol S}$ is referred to as the {\it graph-shift} operator~\cite{sandryhaila2013discrete,shuman2013Emerging}. Possible candidates for ${\boldsymbol S}$  are the graph Laplacian ${\boldsymbol L}$ or the adjacency matrix ${\boldsymbol A}$. Since ${\boldsymbol S}$ is symmetric, it admits the following eigenvalue decomposition
\begin{equation}
\label{eq:graphShift} 
\begin{aligned}
{\boldsymbol S} &= {\boldsymbol U}{\boldsymbol \Lambda}{\boldsymbol U}^H \\
&= [{\boldsymbol u}_1, \cdots, {\boldsymbol u}_N]\, {\rm diag}(\lambda_1,\cdots,\lambda_N) \, [{\boldsymbol u}_1, \cdots, {\boldsymbol u}_N]^H,
\end{aligned}
\end{equation}
where the eigenvectors $\{{\boldsymbol u}_n\}_{n=1}^N$ and the eigenvalues $\{\lambda_n\}_{n=1}^N$ of ${\boldsymbol S}$ provide the notion of frequency in the graph setting~\cite{shuman2013Emerging,sandryhaila2014big}. Specifically, $\{{\boldsymbol u}_n\}_{n=1}^N$ provide a Fourier-like basis for graph signals with the entire spectrum denoted by $\{\lambda_n\}_{n=1}^N$.

The graph shift operator ${\boldsymbol S}$ can be used to define {\it graph filters} of the form~\cite{shuman2013Emerging,sandryhaila2013discrete}
\begin{equation}
\label{eq:graph_fitler}
{\boldsymbol H} = \sum_{l=0}^{L-1} h_l {\boldsymbol S}^l = {\boldsymbol U}\left(\sum_{l=0}^{L-1} h_l{\boldsymbol \Lambda}^l\right){\boldsymbol U}^H,
\end{equation}
where the filter ${\boldsymbol H}$ is of degree $L-1$ with filter coefficients ${\boldsymbol h} = [h_0,h_1,\ldots,h_{L-1}]^T$ and the diagonal matrix 
$\sum_{l=0}^{L-1} h_l{\boldsymbol \Lambda}^l = {\rm diag}({\boldsymbol V}_L{\boldsymbol h})$ can be viewed as the frequency response of the graph filter. Here, ${\boldsymbol V}_L$ is an $N \times L$ Vandermonde matrix with entries $[{\boldsymbol V}]_{i,j} = \lambda_i^{j-1}$. 
%
\vspace*{-2mm}
\subsection{Stationary graph signals}

Let ${\boldsymbol x} = [x_1,x_2,\cdots,x_N]^T \in \mathbb{R}^N$ be a stochastic process defined on the vertices of the graph $\mathcal{G}$ with expected value ${\boldsymbol m} = \mathbb{E}\{{\boldsymbol x}\}$ and covariance matrix ${\boldsymbol R}_{\boldsymbol x} = \mathbb{E}\{({\boldsymbol x} - {\boldsymbol m})({\boldsymbol x} - {\boldsymbol m})^T\}$. The notion of second-order (or wide-sense) stationarity of signals defined over regular structures can be generalized to graph signals as follows. 

\begin{mydef}[Second-order graph stationarity~\cite{benjamin15eusipco,perraudin2016stationary}] 
\vspace*{-2mm}
A stochastic graph process ${\boldsymbol x}$ is graph second-order stationary, if and only if the following properties hold:
\begin{itemize}
\item[1.] The mean of the graph signal is constant, $\mathbb{E}\{x_i\} = m$.
\item[2.] Matrices ${\boldsymbol S}$ and ${\boldsymbol R}_{\boldsymbol x}$ are jointly diagonalizable. 
\end{itemize}
\label{def:graphstationarity}
\end{mydef}
An example of a second-order stationary graph process is white noise with zero mean (thus satisfies the first property) and covariance matrix ${\boldsymbol R}_{\boldsymbol x} = {\boldsymbol I}$, which can be expressed as ${\boldsymbol R}_{\boldsymbol x} = {\boldsymbol U}{\boldsymbol I}{\boldsymbol U}^H$ (thus it can be simultaneously diagonalized with ${\boldsymbol S}$). 

One way to generate second-order stationary graph signals is by (graph) filtering zero-mean unit-variance white noise, which we denote by ${\boldsymbol n} \in \mathbb{R}^N$. In other words, a stochastic graph process ${\boldsymbol x}$ can be modeled as
$
{\boldsymbol x} = {\boldsymbol H}{\boldsymbol n},
$ 
where we recall the graph filter defined in~\eqref{eq:graph_fitler}. It is easy to verify that the filtered signal will have zero mean and covariance matrix ${\boldsymbol R}_{\boldsymbol x} = \mathbb{E}\{({\boldsymbol H}{\boldsymbol n})({\boldsymbol H}{\boldsymbol n})^H\}$ given by
\begin{equation}
\label{eq:cov_filteredsignal}
\begin{aligned}
{\boldsymbol R}_{\boldsymbol x} = {\boldsymbol H}{\boldsymbol H}^H  &= {\boldsymbol U} [{\rm diag}({\boldsymbol V}_L{\boldsymbol h})]^{2} {\boldsymbol U}^H\\
&= {\boldsymbol U} {\rm diag}({\boldsymbol p}) {\boldsymbol U}^H.
\end{aligned}
\end{equation}
This conforms with the second property listed in Definition~\ref{def:graphstationarity}. The diagonal matrix ${\rm diag}({\boldsymbol p})$ is the {\it graph power spectral density} or {\it graph power spectrum} matrix. We formally introduce it through the following definition.

\begin{mydef}[Graph power spectrum]
\vspace*{-2mm}
 The graph power spectral density of a stationary graph process is a real-valued nonnegative length-$N$ vector ${\boldsymbol p}$ defined as
\begin{equation}
\label{eq:graphPSD}
{\rm diag}({\boldsymbol p}) = {\boldsymbol U}^H {\boldsymbol R}_{\boldsymbol x} {\boldsymbol U}.
\end{equation}
Alternatively, $[{\boldsymbol p}]_n = [{\boldsymbol V}_L{\boldsymbol h}]_n^2$.
\end{mydef}
It is worth observing that white noise defined on graphs has a constant graph power spectrum.
In sum, graph stationarity is preserved by linear filtering, thus graph stationary signals with a prescribed graph power spectrum can be generated by filtering white noise. In fact, the graph power spectrum of the filtered signal is reshaped according to the filter. 

\section{Graph power spectrum estimator} \label{sec:GPSDestimation}
The size of the datasets inhibits a direct computation (using a graph Fourier transform matrix) of the graph power spectrum using \eqref{eq:graphPSD} as it requires diagonalization of the graph shift operator that computationally costs $O(N^3)$, and in addition, it requires observing all the nodes for computing ${\boldsymbol R}_{\boldsymbol x}$. In what follows, we introduce the concept of subsampling graph signals for power spectrum estimation, where we leverage the second-order graph stationarity. More specifically, we are interested in determining a reduced set of $K$ graph nodes to sample and in estimating the power spectrum of the entire graph signal from these subsampled observations. This problem is even more challenging in the graph setting as compared to compressive covariance sensing of signals defined over regular structures~\cite{ariananda2012compressive,Romero16CCSspm}. This is  because for signals with regular support, the covariance matrix has some structure (e.g., Toeplitz) that enables elegant subsampling, but for graph signals, the covariance matrix does not admit any known structure, in general.
 %
%

\subsection{Graph spectral domain} \label{sec:GraphSpectralDomain}
Consider the problem of estimating the graph power spectrum of the second-order stationary graph process ${\boldsymbol x} \in \mathbb{R}^N$ from a set of $K \ll N$ linear observations stacked in the vector ${\boldsymbol y} \in \mathbb{R}^K$, given by
\begin{equation}
\label{eq:subsampling}
{\boldsymbol y} = {\boldsymbol \Phi}({\boldsymbol w}){\boldsymbol x} = {\boldsymbol \Phi}({\boldsymbol w}){\boldsymbol H} {\boldsymbol n}. 
\end{equation}
Here, ${\boldsymbol \Phi}({\boldsymbol w}) = {\rm diag_r}({\boldsymbol w}) \in \{0,1\}^{K \times N}$ is a {\it sparse sampling} or {\it subsampling} matrix guided by a {\it component selection} vector  ${\boldsymbol w} = [w_1,\cdots,w_N]^T\in \{0,1\}^N$, where $w_i=1$ indicates that the $i$th graph node is selected, otherwise it is not selected (${\rm diag_r}(\cdot)$ represents a diagonal matrix with the argument on its diagonal but with the all-zero rows removed).

Using the subsampling scheme in \eqref{eq:subsampling}, the covariance matrix of the subsampled graph process ${\boldsymbol y}$ can be computed as 
\begin{equation}
\label{eq:subsamCov}
{\boldsymbol R}_{\boldsymbol y} = {\boldsymbol \Phi}{\boldsymbol R}_{\boldsymbol x}{\boldsymbol \Phi}^H = {\boldsymbol \Phi}{\boldsymbol U} {\rm diag}({\boldsymbol p}) {\boldsymbol U}^H{\boldsymbol \Phi}^H \in \mathbb{R}^{K \times K},
\end{equation}
where we simply write ${\boldsymbol \Phi}({\boldsymbol w})$ as ${\boldsymbol \Phi}$ for conciseness. Vectorizing\footnote{We use the matrix property ${\rm vec}({\boldsymbol A}{\rm diag}({\boldsymbol d}){\boldsymbol B})=({\boldsymbol B}^H \circ {\boldsymbol A}){\boldsymbol d}$, where  $\circ$ denotes the Khatri-Rao or columnwise Kronecker product, $\otimes$ denotes the Kronecker product, and ${\rm vec}(\cdot)$ is the matrix vectorization operator.}~\eqref{eq:subsamCov}, we obtain a set of $K^2$ equations in $N$ unknowns:
\begin{equation}
\label{eq:subsamCovVec}
\begin{aligned}
{\boldsymbol r}_{\boldsymbol y} = {\rm vec}({\boldsymbol R}_{\boldsymbol y})  &=({\boldsymbol \Phi}{\boldsymbol U} \circ {\boldsymbol \Phi}{\boldsymbol U}) {\boldsymbol p} \\
&\overset{(a)}{=}({\boldsymbol \Phi} \otimes {\boldsymbol \Phi}) ({\boldsymbol U} \circ  {\boldsymbol U}){\boldsymbol p} 
{=}({\boldsymbol \Phi} \otimes {\boldsymbol \Phi}) {\boldsymbol \Psi}_{\rm s}{\boldsymbol p},
\end{aligned}
\end{equation}
where $(a)$ is due to the matrix property $({\boldsymbol A} \otimes {\boldsymbol B}) ({\boldsymbol C} \circ  {\boldsymbol D})  = ({\boldsymbol A}{\boldsymbol C} \circ {\boldsymbol B}{\boldsymbol D})$, and the subscript ``${\rm s}$" in ${\boldsymbol \Psi}_{\rm s}$ (constructed using the graph Fourier matrix) stands for {\it spectral domain} approach. If the matrix $({\boldsymbol \Phi}{\boldsymbol U} \circ {\boldsymbol \Phi}{\boldsymbol U})$ has full column rank, which requires $K^2 \geq N$, then the graph power spectrum can be estimated in closed form via least squares, given by 
\[
\widehat{\boldsymbol p} = ({\boldsymbol \Phi}{\boldsymbol U} \circ {\boldsymbol \Phi}{\boldsymbol U})^\dag {\boldsymbol r}_{\boldsymbol y},
\]
where for a full column rank matrix ${\boldsymbol A}$, we have ${\boldsymbol A}^\dag = ({\boldsymbol A}^T{\boldsymbol A})^{-1}{\boldsymbol A}^T$.
\begin{myrem}[Spectral priors] A higher compression can be achieved if we have some prior knowledge about the graph spectra. More specifically, it is possible to have $K^2 < N$, if we know a priori that (a) the spectrum is bandlimited (e.g., lowpass) with known support, or (b) the spectrum is sparse, but with unknown support. Further, this information can be included while estimating the graph power spectrum from \eqref{eq:subsamCovVec}, e.g., using a reduced-order least squares or an $\ell_1$-norm regularized least squares.
\end{myrem}

\subsection{Graph vertex domain} \label{sec:GraphVertexDomain}
The covariance matrix of a stochastic graph process and the graph shift operator can be simultaneously diagonalized. This allows us to express the covariance matrix ${\boldsymbol R}_{\boldsymbol x}$ as a polynomial of the graph shift operator of the form:
\begin{equation}
\label{eq:Rxpolynom}
{\boldsymbol R}_{\boldsymbol x} =  \sum_{q=0}^{Q-1} \alpha_q {\boldsymbol S}^q,
\end{equation}
where the $Q=\min\{2L-1,N\}$ unknown expansion coefficients $\{\alpha_q\}_{q=0}^{Q-1}$ collected in the vector ${\boldsymbol \alpha} = [\alpha_0,\alpha_1,\cdots,\alpha_{Q-1}]^T \in \mathbb{R}^Q$ completely characterize the covariance matrix. In other words, we assume a 
linear parametrization of the covariance matrix ${\boldsymbol R}_{\boldsymbol x}$ using the set of $Q$ symmetric matrices $\{{\boldsymbol S}^0,{\boldsymbol S},\cdots,{\boldsymbol S}^{Q-1}\} \subset \mathbb{S}^N$ as a basis. 

Vectorizing ${\boldsymbol R}_{\boldsymbol x}$ in \eqref{eq:Rxpolynom} yields
\begin{equation}
\label{eq:RxpolynomVec}
{\boldsymbol r}_{\boldsymbol x} = {\rm vec}({\boldsymbol R}_{\boldsymbol x}) = \sum_{q=0}^{Q-1} \alpha_q {\rm vec}({\boldsymbol S}^q) = {\boldsymbol \Psi}_{\rm v}{\boldsymbol \alpha},
\end{equation}
where we have stacked ${\rm vec}({\boldsymbol S}^q)$ to form columns of the matrix ${\boldsymbol \Psi}_{\rm v} \in \mathbb{R}^{N^2 \times Q}$, and the subscript ``${\rm v}$" in ${\boldsymbol \Psi}_{\rm v}$ stands for {\it vertex domain} approach.
Using the matrix property  ${\rm vec}({\boldsymbol A}{\boldsymbol B}{\boldsymbol C})=({\boldsymbol C}^H \otimes {\boldsymbol A}){\rm vec}({\boldsymbol B})$, the covariance matrix of the subsampled graph process [cf.~\eqref{eq:subsamCov}] ${\boldsymbol R}_{\boldsymbol y} =  {\boldsymbol \Phi}{\boldsymbol R}_{\boldsymbol x}{\boldsymbol \Phi}^H$, can be vectorized to obtain a set of $K^2$ equations in $Q$ unknowns, given by 
\begin{equation}
\label{eq:ry_vertex}
\begin{aligned}
{\boldsymbol r}_{\boldsymbol y} = {\rm vec}({\boldsymbol R}_{\boldsymbol y}) &= ({\boldsymbol \Phi} \otimes {\boldsymbol \Phi}) {\rm vec}({\boldsymbol R}_{\boldsymbol x})\\
&= ({\boldsymbol \Phi} \otimes {\boldsymbol \Phi}){\boldsymbol \Psi}_{\rm v}{\boldsymbol \alpha}. 
\end{aligned}
\end{equation}
If the matrix $({\boldsymbol \Phi} \otimes {\boldsymbol \Phi}){\boldsymbol \Psi}_{\rm v}$ has full column rank, which requires $K^2 \geq Q$, then the overdetermined system can be solved using least squares as
\[
\widehat{\boldsymbol \alpha} = [({\boldsymbol \Phi} \otimes {\boldsymbol \Phi}){\boldsymbol \Psi}_{\rm v}]^\dag {\boldsymbol r}_{\boldsymbol y}.
\] 
The problem of estimating $\{\alpha_q\}_{q=0}^{Q-1}$ is known as covariance matching~\cite{ottersten1998covariance}, and in the graph setting we refer to it as {\it graph covariance matching}. The computationally expensive eigenvalue decomposition of the graph Laplacian that costs $O(N^3)$ is not needed to reconstruct~${\boldsymbol \alpha}$.
%

The graph power spectrum can be subsequently recovered according to the following remark.
\begin{myrem}
We can relate the vector ${\boldsymbol p}$ and the vector ${\boldsymbol \alpha}$, by using \eqref{eq:graphPSD} and \eqref{eq:Rxpolynom}. That is, we can write ${\rm diag}({\boldsymbol p}) = \sum_{q=0}^{Q-1} \alpha_q {\boldsymbol \Lambda}^q$, or in matrix-vector form we have 
${\boldsymbol p} = {\boldsymbol V}_Q {\boldsymbol \alpha}$, where ${\boldsymbol V}_Q$ is an $N \times Q$ Vandermonde matrix with entries $ [{\boldsymbol V}_Q]_{i,j} = \lambda_i^{j-1}$. To recover ${\boldsymbol p}$ from ${\boldsymbol \alpha}$, however, requires all the $N$ eigenvalues to construct ${\boldsymbol V}_Q$. 
\end{myrem}

\begin{algorithm}[!b]
\caption{Greedy algorithm}
\label{alg:greedy}
\begin{algorithmic}
\item[1.] \textbf{Require} $\mathcal{X} = \emptyset, K$.
\item[2.] \textbf{for} $k=1$ to $K$
\item[3.] \hskip1cm$
s^\ast = \argmax\limits_{ s \notin \mathcal{X}}   \, f (\mathcal{X} \cup \{s\}) 
$
\item[4.] \hskip1cm$
\mathcal{X} \leftarrow \mathcal{X} \cup \{s^\ast\}
$
\item[5.] \textbf{end} 
\item[6.] \textbf{Return} $\mathcal{X}$
\end{algorithmic}
\end{algorithm} 

Finally, we show the equivalence between linear models \eqref{eq:subsamCovVec} and \eqref{eq:ry_vertex} as follows.
\begin{myrem}
The fact that ${\boldsymbol S}^q = {\boldsymbol U}{\boldsymbol \Lambda}^q{\boldsymbol U}^H$ from \eqref{eq:graphShift} allows us to express 
${\boldsymbol \Psi}_{\rm v}$ in \eqref{eq:ry_vertex} as
\begin{equation}
\label{eq:PsivecS}
{\boldsymbol \Psi}_{\rm v} = ({\boldsymbol U} \circ  {\boldsymbol U}){\boldsymbol V}_Q.
\end{equation}
Using \eqref{eq:PsivecS} in \eqref{eq:ry_vertex}, we get 
\begin{equation}
\label{eq:ry_vertexSim}
\begin{aligned}
{\boldsymbol r}_{\boldsymbol y} = ({\boldsymbol \Phi} \otimes {\boldsymbol \Phi}) ({\boldsymbol U} \circ  {\boldsymbol U}) {\boldsymbol V}_Q {\boldsymbol \alpha} = ({\boldsymbol \Phi}{\boldsymbol U} \circ {\boldsymbol \Phi}{\boldsymbol U})  {\boldsymbol p}.
\end{aligned}
\end{equation}
\end{myrem}
%
\begin{figure*}[!t]
\psfrag{random graph}{}
\psfrag{random graph vertex domain}{}
\psfrag{Laplacian eigenvalues}{\footnotesize Laplacian eigenvalues}
\psfrag{Modeled PSD (uncompressed)}{ \footnotesize Estimated PSD ($0\%$ compression)}
\psfrag{Real PSD}{ \footnotesize True PSD}
        \centering
        	\begin{subfigure}[b]{0.45\textwidth}
                \centering
                \includegraphics[width=\columnwidth, height=1.5in]{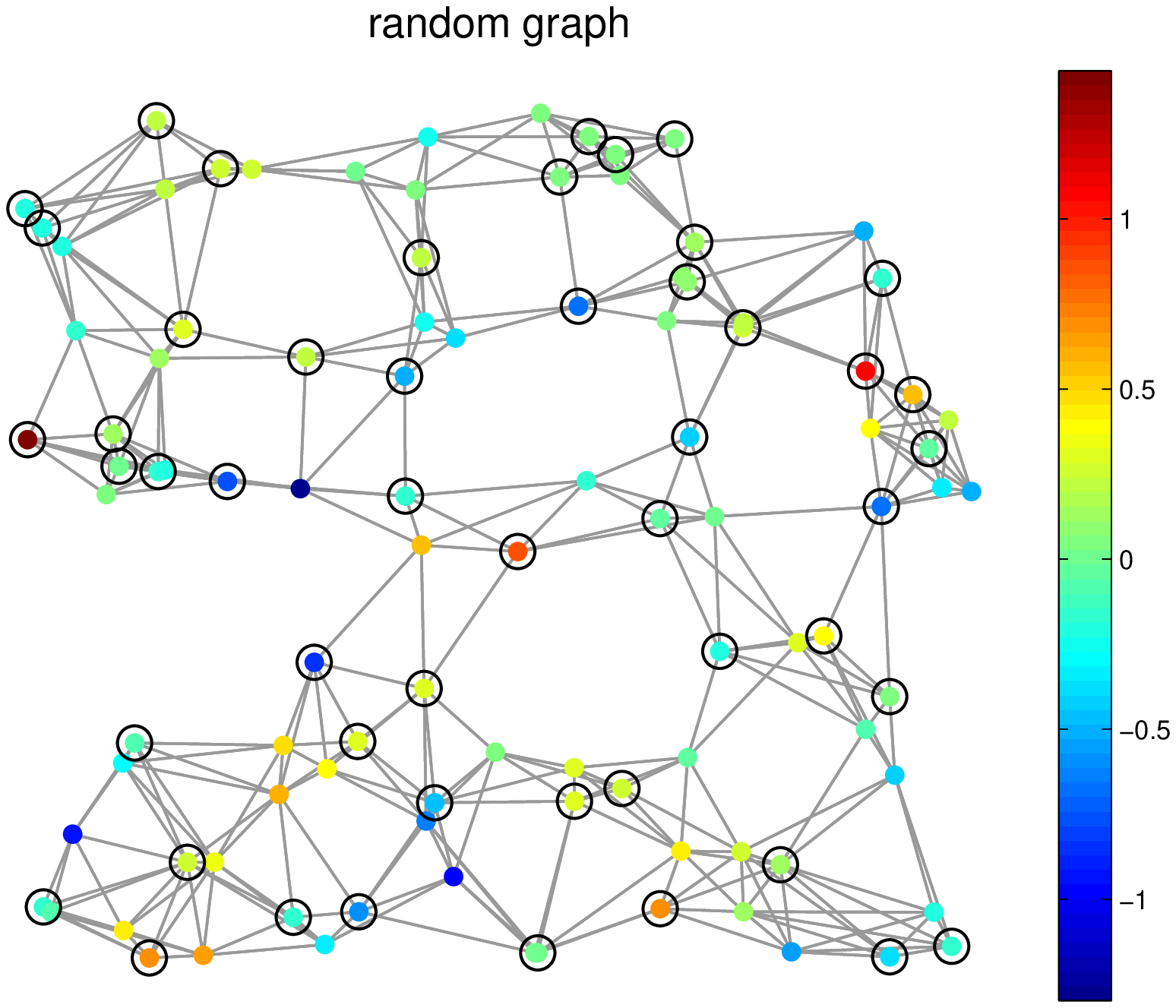}
                \label{fig:spectral_graphfig}
        \end{subfigure} 
~
       \begin{subfigure}[b]{0.45\textwidth}
       \psfrag{Estimated PSD (uncompressed)}{\footnotesize Estimated PSD ($0\%$ compression)}
              \psfrag{Estimated PSD (compressed)}{\footnotesize Estimated PSD ($50\%$ compression)}
 	\centering
                \includegraphics[width=\columnwidth, height=1.5in]{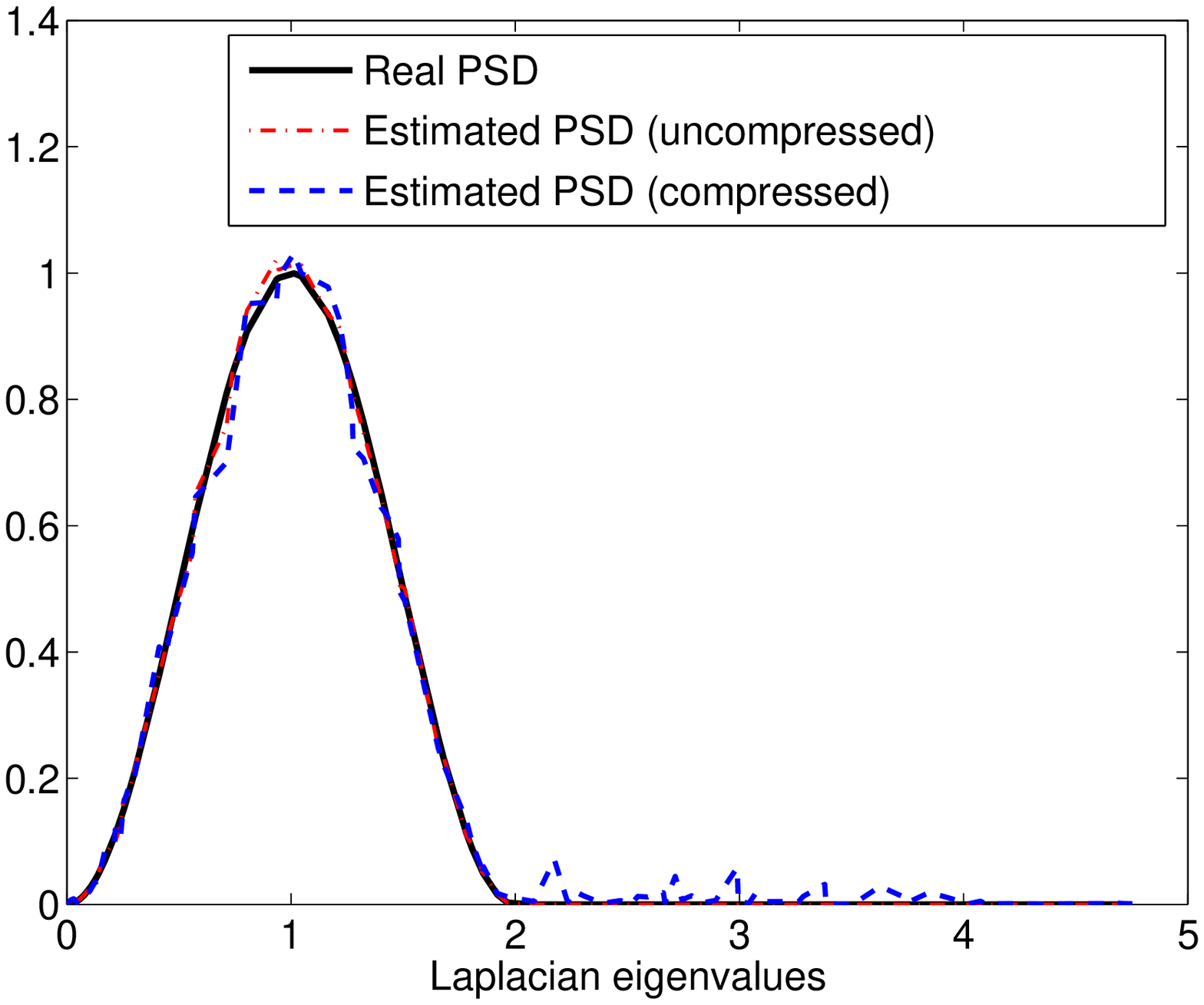}
                \label{fig:spectral_psdfig}
        \end{subfigure}%
                        \vspace*{-5mm}
        \caption{\footnotesize{Graph spectral domain subsampling. Left: Random graph with $N=100$ nodes. Sampled $K=50$ nodes are depicted with a black circle. Right: True and estimated graph power spectrum.}}\label{fig:spectral}

\psfrag{random graph}{}
\psfrag{random graph vertex domain}{}
\psfrag{Laplacian eigenvalues}{\footnotesize Laplacian eigenvalues}
\psfrag{Modeled PSD (uncompressed)}{ \footnotesize Estimated PSD ($0\%$ compression)}
\psfrag{Real PSD}{ \footnotesize True PSD}
        \centering

       \begin{subfigure}[b]{0.45\textwidth}
 	\centering
                \includegraphics[width=\columnwidth, height=1.5in]{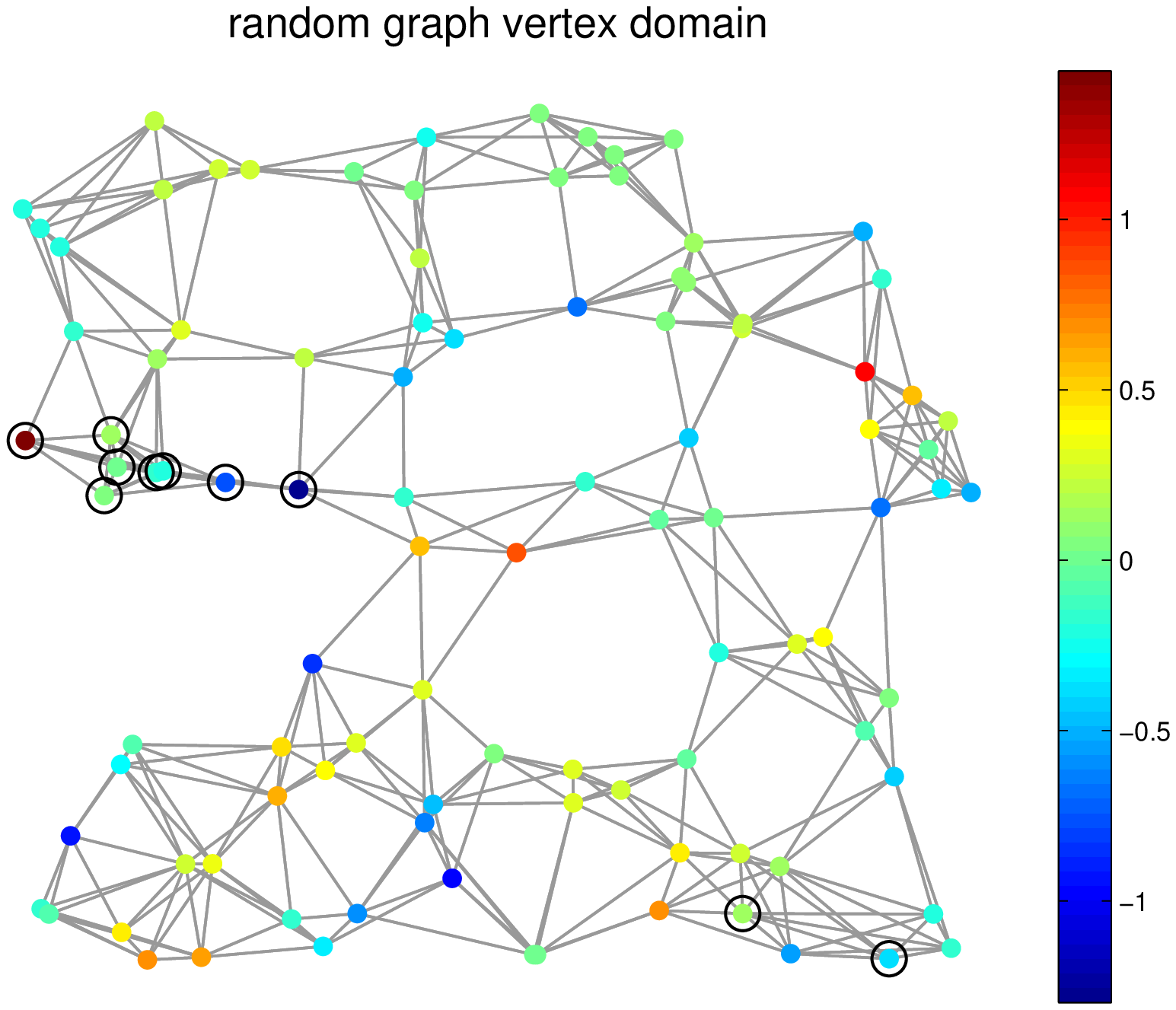}
                \label{fig:vertex_graphfig}
        \end{subfigure}%
~
       \begin{subfigure}[b]{0.45\textwidth}
       \psfrag{Estimated PSD (compressed)}{\footnotesize Estimated PSD ($90\%$ compression)}
 	\centering
                \includegraphics[width=\columnwidth, height=1.5in]{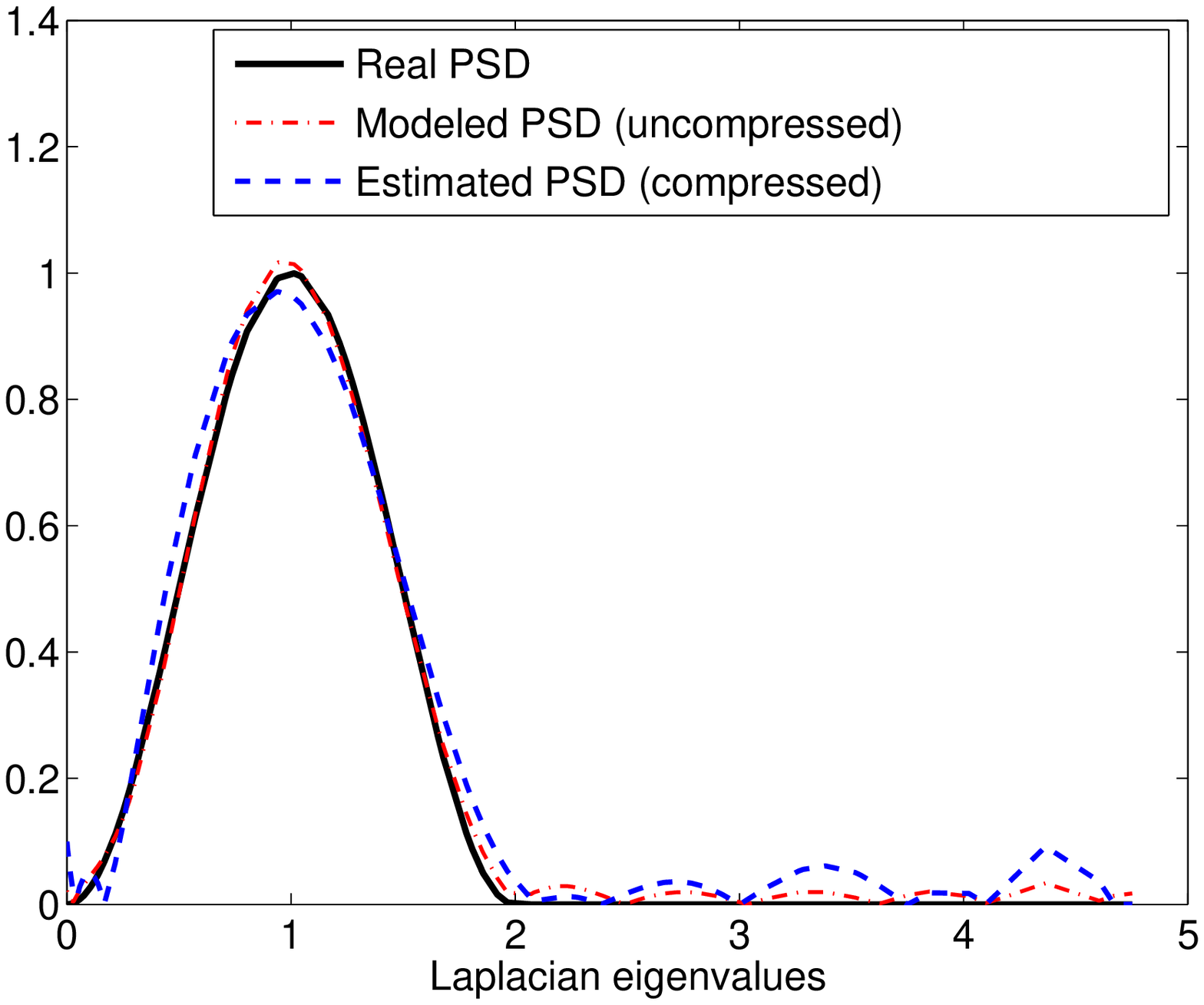}
                \label{fig:vertex_psdfig}
        \end{subfigure}%
                              \vspace*{-5mm}           
        \caption{\footnotesize{Graph vertex domain subsampling. Left: Random graph with $N=100$ nodes. Sampled $K=10$ nodes are depicted with a black circle. Right: Graph power spectrum, where the true power spectrum is modeled with $Q=12$.}}\label{fig:vertex}
            \vspace*{-4mm}
\end{figure*}

\section{Subsampler design}

If we can design a full-column rank model matrix $({\boldsymbol \Phi} \otimes {\boldsymbol \Phi}){\boldsymbol \Psi}$ with ${\boldsymbol \Psi}:={\boldsymbol \Psi}_{\rm s}$ or  ${\boldsymbol \Psi}:={\boldsymbol \Psi}_{\rm v}$, then we can perfectly recover the graph power spectrum by observing a reduced set of only $K$ graph nodes. We will develop a low-complexity algorithm to design such full-column rank matrices in this section.

The least squares solution developed in \S \ref{sec:GPSDestimation} depends on the spectrum of  
\begin{align*}
{\boldsymbol T}({\boldsymbol w}) = [({\boldsymbol \Phi}({\boldsymbol w}) \otimes {\boldsymbol \Phi}({\boldsymbol w})){\boldsymbol \Psi}({\boldsymbol w})]^T[({\boldsymbol \Phi}({\boldsymbol w}) \otimes {\boldsymbol \Phi}({\boldsymbol w})){\boldsymbol \Psi}]
\end{align*}
i.e., the performance of least squares is better if the spectrum of $({\boldsymbol \Phi} \otimes {\boldsymbol \Phi}){\boldsymbol \Psi}$ is more uniform. Thus a good sparse sampler ${\boldsymbol w}$ can be obtained by solving:
\begin{equation}
\label{eq:logdetopt}
\argmax_{{\boldsymbol w} \in \{0,1\}^N} \quad f({\boldsymbol w}) \quad{\rm s.t.}\quad  \|{\boldsymbol w}\|_0 = K
\end{equation}
with either $f({\boldsymbol w}) = \lambda_{\rm min}\{{\boldsymbol T}({\boldsymbol w})\}$ or $f({\boldsymbol w}) = \log\det \{{\boldsymbol T}({\boldsymbol w})\}$, both of which try of balance the spectrum of ${\boldsymbol T}({\boldsymbol w})$. Although the above Boolean nonconvex problem with $f({\boldsymbol w}) = \lambda_{\rm min}\{{\boldsymbol T}({\boldsymbol w})\}$ and $f({\boldsymbol w}) = \log\det \{{\boldsymbol T}({\boldsymbol w})\}$ can be relaxed and solved using convex optimization (e.g., see~\cite{ChepuriTSPsel, chepuri2014spl}), we will focus on the optimization problem \eqref{eq:logdetopt} with $f({\boldsymbol w}) = \log\det \{{\boldsymbol T}({\boldsymbol w})\}$ as it can be solved near-optimally using a low-complexity greedy algorithm.

Let us define an index set $\mathcal{X}$ that is related to the component selection vector ${\boldsymbol w}$ as 
$
\mathcal{X} = \{ m \, | \, w_m=1,m=1,\ldots,N\},
$ 
where $\mathcal{X} \subseteq \mathcal{N}$ with $\mathcal{N} = \{1,\ldots,N\}$.
We can now express the cost function $f({\boldsymbol w}) = \log\det \{{\boldsymbol T}({\boldsymbol w})\}$ equivalently as the set function 
given by 
\begin{equation}
\label{eq:submodularlogdet}
f(\mathcal{X}) = \log \det \left\{\sum\nolimits_{{(i,j)} \in \mathcal{X} \times \mathcal{X}} {\boldsymbol \psi}_{i,j} {\boldsymbol \psi}_{i,j}^T\right\},
\end{equation}
where the $N^2$ column vectors $\{{\boldsymbol \psi}_{1,1},{\boldsymbol \psi}_{1,2}, \cdots,{\boldsymbol \psi}_{N,N}\}$ are used to form the rows of ${\boldsymbol \Psi}$ as ${\boldsymbol \Psi} = [{\boldsymbol \psi}_{1,1},{\boldsymbol \psi}_{1,2}, \cdots,{\boldsymbol \psi}_{N,N}]^T$. We use such an indexing because the sampling matrix ${\boldsymbol \Phi} \otimes {\boldsymbol \Phi}$ results in a structured (row) subset selection. 

{\it Submodularity} \textemdash a notion based on the property of diminishing returns, is useful for solving discrete combinatorial optimization problems of the form \eqref{eq:logdetopt} (see e.g., \cite{krause2008optimizing}).  Submodularity can be formally defined as follows.

\begin{mydef}[Submodular function]
Given two sets $\mathcal{X}$ and $\mathcal{Y}$ such that for every $\mathcal{X} \subseteq \mathcal{Y} \subseteq \mathcal{N}$ and $s \in \mathcal{M} \backslash \mathcal{Y}$, the set function $f: 2^\mathcal{N} \rightarrow \mathbb{R}$ defined on the subsets of $\mathcal{N}$ is said to be submodular, if it satisfies
\[
f(\mathcal{X} \cup \{s\}) - f(\mathcal{X}) \geq f(\mathcal{Y} \cup \{s\}) - f(\mathcal{Y}).
\]
\end{mydef}
Further, if the submodular function is monotone nondecreasing, that is, $f(\mathcal{X})$ $\leq f(\mathcal{Y})$ for all $\mathcal{X} \subseteq \mathcal{Y} \subseteq \mathcal{N}$ and normalized (i.e., $f(\emptyset) = 0$), then a greedy maximization of such a function via Algorithm~\ref{alg:greedy} is near optimal with an approximation factor of $(1 -1/e)$, where $e$ is Euler's number \cite{nemhauser1978analysis}. That is, $f(\mathcal{X}) \geq (1 -1/e) f({\rm OPT}),$ where  $f({\rm OPT}) = \max_{\mathcal{X} \subset \mathcal{N}, |\mathcal{X}| = K} \, f(\mathcal{X})$.
The cost function \eqref{eq:submodularlogdet} after a slight modification satisfies the above property as stated in the following lemma.
\begin{mylem} \label{lem:submodular_logdet}
The set function $f: 2^\mathcal{N} \rightarrow \mathbb{R}$ given by
\begin{align}
\label{eq:submodularlogdet_normalized}
\hskip-2mmf(\mathcal{X}) = \log \det \left\{\sum\nolimits_{{(i,j)} \in \mathcal{X} \times \mathcal{X}} {\boldsymbol \psi}_{i,j} {\boldsymbol \psi}_{i,j}^T + \epsilon {\boldsymbol I}\right\} - N \log \epsilon
\end{align}
is a normalized, nonnegative monotone, submodular function on the set $\mathcal{X} \subset \mathcal{N}$. Here, $\epsilon >0$ is a small constant. Hence \eqref{eq:submodularlogdet_normalized} is a reasonable approximation of \eqref{eq:submodularlogdet}.  
\end{mylem}

In \eqref{eq:submodularlogdet_normalized}, $N \log \epsilon$ ensures that $f(\emptyset)$ is zero. Using the result from~\cite{shamaiah2010greedy} that the set function $g: 2^\mathcal{N} \rightarrow \mathbb{R}$, given by 
\begin{align}
\label{eq:classicallogdet}
g(\mathcal{X}) = \log \det \left\{\sum\nolimits_{i \in \mathcal{X}} {\boldsymbol a}_{i} {\boldsymbol a}_{i}^T + \epsilon {\boldsymbol I}\right\} 
- N \log \epsilon
\end{align}
with column vectors $\{{\boldsymbol a}_{i}\}_{i=1}^N$ is a normalized, nonnegative monotone, submodular function on the set $\mathcal{X} \subset \mathcal{N}$, we can prove Lemma~\ref{lem:submodular_logdet}. Therefore, the solution based on the greedy algorithm summarized as Algorithm~\ref{alg:greedy} results in a $(1-1/e)$ optimal solution for~\eqref{eq:logdetopt}. Note that the number of summands 
in \eqref{eq:classicallogdet} and \eqref{eq:submodularlogdet_normalized}, is respectively, $|\mathcal{X}|$ and $|\mathcal{X}|^2$.
It is worth mentioning that the greedy algorithm is linear in $K$, while computing \eqref{eq:submodularlogdet_normalized} remains the dominating cost. Nevertheless, \eqref{eq:submodularlogdet_normalized} can be computed efficiently using rank-1 updates, similar to \cite{shamaiah2010greedy}. 

Other submodular functions that promote full-column rank model matrices, e.g., frame potential~\cite{vertterlisensorplacement} defined as $f({\boldsymbol w}) = {\rm tr}\{{\boldsymbol T}^H({\boldsymbol w}){\boldsymbol T}({\boldsymbol w})\}$, are also reasonable costs to optimize. 
Finally, random subsampling is not suitable as it might not always result in a full-column rank model matrix.
\vspace*{-1mm}
\section{Numerical experiments}

In this section we test the practical performance of the proposed estimator as well the designed sparse sampler. For the experiments, we use a random sensor graph with $N=100$ nodes generated using the {\tt GSPBOX}~\cite{perraudin2014gspbox}. The graph topology can be seen on the left side of Figure~\ref{fig:spectral} and Figure~\ref{fig:vertex} along with a random signal realization. Graph stationary signals are generated by filtering zero-mean unit-variance white noise with a lowpass filter, which has a squared magnitude frequency response as shown in Figure~\ref{fig:spectral} (labeled as ``True PSD") and it has $L=7$ filter coefficients. We use $N_s = 1000$ snapshots to form a sample covariance matrix, which we use in the experiments.

In the graph spectral domain approach, using Algorithm~\ref{alg:greedy}, we first design the subsampler by selecting rows of the matrix ${\boldsymbol \Psi}_{\rm s}$ in a structured manner determined by ${\boldsymbol \Phi} \otimes {\boldsymbol \Phi}$, one by one. In other words, we perform a row subset selection of the (modified) graph Fourier matrix ${\boldsymbol U} \circ {\boldsymbol U}$. For this particular scenario, a full-column rank matrix $({\boldsymbol \Phi} \otimes {\boldsymbol \Phi}){\boldsymbol \Psi}_{\rm s}$ was obtained for $K > 11$. We show on the right side of Figure~\ref{fig:spectral}, the reconstructed graph power spectrum for $K=50$ (i.e., $50\%$ compression) as well as for $K=N$ (it is the diagonal of the sample covariance matrix with no compression). On the left side of Figure~\ref{fig:spectral}, we show the selected graph nodes with a black circle. 

In the graph vertex domain approach, we use $Q=12$ to construct the model matrix ${\boldsymbol \Psi}_{\rm v}$. As before, we perform a row subset selection of the matrix ${\boldsymbol \Psi}_{\rm s}$ in a structured way using Algorithm~\ref{alg:greedy}. We show on the right side of Figure~\ref{fig:vertex}, the least squares estimate of the graph power spectrum using $K=10$ (i.e., a compression of $90\%$). Such a high compression is possible because we a priori know the low value of $Q$.

\vspace*{-1.5mm}
\section{Conclusions}

In this paper we have investigated sampling of random processes defined on graphs. In particular, we have focused on subsampling stationary graph signals for estimating the power spectral density. We have shown that it is possible to subsample as low as $O(\sqrt{N})$ vertices and yet reconstruct the power spectrum of a signal defined on a graph with $N$ vertices, without any spectral priors. The subsamplers are designed using a greedy algorithm, which near optimally solves the combinatorial Boolean optimization problem. A least squares estimator has been proposed to reconstruct the graph power spectrum from the subsampled observations.


\vfill
\newpage
%
\bibliographystyle{IEEEtran}
\bibliography{IEEEabrv,//users/localadmin/Dropbox/Bibfiles/refs,//users/localadmin/Dropbox/Bibfiles/strings}

\begin{thebibliography}{10}
\providecommand{\url}[1]{#1}
\csname url@samestyle\endcsname
\providecommand{\newblock}{\relax}
\providecommand{\bibinfo}[2]{#2}
\providecommand{\BIBentrySTDinterwordspacing}{\spaceskip=0pt\relax}
\providecommand{\BIBentryALTinterwordstretchfactor}{4}
\providecommand{\BIBentryALTinterwordspacing}{\spaceskip=\fontdimen2\font plus
\BIBentryALTinterwordstretchfactor\fontdimen3\font minus
  \fontdimen4\font\relax}
\providecommand{\BIBforeignlanguage}[2]{{%
\expandafter\ifx\csname l@#1\endcsname\relax
\typeout{** WARNING: IEEEtran.bst: No hyphenation pattern has been}%
\typeout{** loaded for the language `#1'. Using the pattern for}%
\typeout{** the default language instead.}%
\else
\language=\csname l@#1\endcsname
\fi
#2}}
\providecommand{\BIBdecl}{\relax}
\BIBdecl

\bibitem{shuman2013Emerging}
D.~I. Shuman, S.~K. Narang, P.~Frossard, A.~Ortega, and P.~Vandergheynst, ``The
  emerging field of signal processing on graphs: Extending high-dimensional
  data analysis to networks and other irregular domains,'' \emph{{IEEE} Signal
  Process. Mag.}, vol.~30, no.~3, pp. 83--98, 2013.

\bibitem{sandryhaila2014big}
A.~Sandryhaila and J.~M. Moura, ``Big data analysis with signal processing on
  graphs: Representation and processing of massive data sets with irregular
  structure,'' \emph{{IEEE} Signal Process. Mag.}, vol.~31, no.~5, pp. 80--90,
  2014.

\bibitem{girault2014semi}
B.~Girault, P.~Goncalves, E.~Fleury, and A.~S. Mor, ``Semi-supervised learning
  for graph to signal mapping: a graph signal wiener filter interpretation,''
  in \emph{Proc. of IEEE International Conference on Acoustics, Speech and
  Signal Processing}, May 2014, Florence, Italy, 2014.

\bibitem{anis2014towards}
A.~Anis, A.~Gadde, and A.~Ortega, ``Towards a sampling theorem for signals on
  arbitrary graphs,'' in \emph{Proc. of IEEE International Conference on
  Acoustics, Speech and Signal Processing}, May 2014, Florence, Italy, 2014.

\bibitem{marques2015sampling}
A.~G. Marques, S.~Segarra, G.~Leus, and A.~Ribeiro, ``Sampling of graph signals
  with successive local aggregations,'' \emph{{IEEE} Trans. Signal Process.},
  vol.~64, no.~7, pp. 1832--1843, 2016.

\bibitem{tsitsvero2015uncertainty}
M.~Tsitsvero, S.~Barbarossa, and P.~D. Lorenzo, ``Uncertainty principle and
  sampling of signals defined on graphs,'' in \emph{Proc. of 49th Asilomar
  Conference on Signals, Systems and Computers}, Nov. 2014, California, USA,
  2015.

\bibitem{ariananda2012compressive}
D.~D. Ariananda and G.~Leus, ``Compressive wideband power spectrum
  estimation,'' \emph{{IEEE} Trans. Signal Process.}, vol.~60, no.~9, pp.
  4775--4789, 2012.

\bibitem{Romero16CCSspm}
D.~Romero, D.~D. Ariananda, Z.~Tian, and G.~Leus, ``Compressive covariance
  sensing: Structure-based compressive sensing beyond sparsity,'' \emph{{IEEE}
  Signal Process. Mag.}, vol.~33, no.~1, pp. 78--93, Jan 2016.

\bibitem{sandryhaila2013discrete}
A.~Sandryhaila and J.~M. Moura, ``Discrete signal processing on graphs,''
  \emph{{IEEE} Trans. Signal Process.}, vol.~61, no.~7, pp. 1644--1656, 2013.

\bibitem{benjamin15eusipco}
B.~Girault, ``Stationary graph signals using an isometric graph translation,''
  in \emph{Proc. of 23rd European Signal Processing Conference}, Aug 2015,
  Nice, France, 2015.

\bibitem{perraudin2016stationary}
N.~Perraudin and P.~Vandergheynst, ``Stationary signal processing on graphs,''
  \emph{arXiv preprint arXiv:1601.02522}, 2016.

\bibitem{ottersten1998covariance}
B.~Ottersten, P.~Stoica, and R.~Roy, ``Covariance matching estimation
  techniques for array signal processing applications,'' \emph{Digital Signal
  Processing}, vol.~8, no.~3, pp. 185--210, 1998.

\bibitem{ChepuriTSPsel}
S.~P. Chepuri and G.~Leus, ``Sparsity-promoting sensor selection for non-linear
  measurement models,'' \emph{{IEEE} Trans. Signal Process.}, vol.~63, no.~3,
  pp. 684--698, Feb. 2015.

\bibitem{chepuri2014spl}
------, ``Continuous sensor placement,'' \emph{{IEEE} Signal Process. Lett.},
  vol.~22, no.~5, pp. 544--548, May 2015.

\bibitem{krause2008optimizing}
A.~Krause, \emph{Optimizing sensing: Theory and applications}, ser. {Ph.D.}
  dissertation, School of Comput. Sci.\hskip 1em plus 0.5em minus 0.4em\relax
  Carnegie Mellon Univ., Pittsburgh, PA, United States, 2008.

\bibitem{nemhauser1978analysis}
G.~L. Nemhauser, L.~A. Wolsey, and M.~L. Fisher, ``An analysis of
  approximations for maximizing submodular set functions{\textemdash} {I},''
  \emph{Mathematical Programming}, vol.~14, no.~1, pp. 265--294, 1978.

\bibitem{shamaiah2010greedy}
M.~Shamaiah, S.~Banerjee, and H.~Vikalo, ``Greedy sensor selection: Leveraging
  submodularity,'' in \emph{Proc. of 49th IEEE Conference on Decision and
  Control}, Dec. 2010, Atlanta, Georgia, USA, 2010.

\bibitem{vertterlisensorplacement}
J.~Ranieri, A.~Chebira, and M.~Vetterli, ``Near-optimal sensor placement for
  linear inverse problems,'' \emph{{IEEE} Trans. Signal Process.}, vol.~62,
  no.~5, pp. 1135--1146, Mar. 2014.

\bibitem{perraudin2014gspbox}
N.~{Perraudin}, J.~{Paratte}, D.~{Shuman}, V.~{Kalofolias}, P.~{Vandergheynst},
  and D.~K. {Hammond}, ``{GSPBOX: A toolbox for signal processing on graphs},''
  \emph{ArXiv e-prints}, Aug. 2014.

\end{thebibliography}

\end{document}